\documentclass[amsmath,amssymb,nofootinbib,onecolumn,a4paper]{revtex4}
\usepackage{graphicx}
\usepackage{bm}

\newcommand{\be}{\begin{equation}}
\newcommand{\ee}{\end{equation}}
\newcommand{\bea}{\begin{eqnarray}}
\newcommand{\eea}{\end{eqnarray}}
\newcommand{\bdm}{\begin{displaymath}}
\newcommand{\edm}{\end{displaymath}}
\newcommand{\beas}{\begin{eqnarray*}}
\newcommand{\eeas}{\end{eqnarray*}}

\newcommand{\mH}{\mathcal{H}}
\newcommand{\mHV}{\tilde{\mathcal{H}}}
\newcommand{\bkr}{\overline{\rho}}
\newcommand{\bkp}{\overline{p}}

\newcommand{\kv}{\mathbf{k}}
\newcommand{\xv}{\mathbf{x}}

\newcommand{\dtk}{\frac{d^3\kv}{(2\pi)^3}}

\def\X{{\cal{X}}}
\def\Xv{{{\cal{X}}_{\rm{v}}}}
\def\XB{{{\cal{X}}_{\rm{B}}}}

\newcommand{\av}[1]{\left<#1\right>}
\newcommand{\pd}{\partial}

\begin{document}

\title{On the physical effects of consistent cosmological averaging}
\author{Iain A. Brown$^{1,2}$}
\email{ibrown@astro.uio.no}
\author{Alan A. Coley$^2$}
\email{aac@mathstat.dal.ca}
\author{D. Leigh Herman$^{2,3}$}
\email{dherman@grenfell.mun.ca}
\author{Joey Latta$^2$}
\email{lattaj@mathstat.dal.ca}
\affiliation{${}^1$Institute of Theoretical Astrophysics, University of Oslo, P.O. Box 6094, N-0315 Blindern, Norway\\${}^2$Department of Mathematics and Statistics, Dalhousie University, 6316 Coburg Road, PO Box 15000, Halifax, Nova Scotia, B3H 4R2, Canada\\${}^3$Division of Science, Memorial University, Grenfell Campus, 20 University Drive, Corner Brook, Newfoundland, A2H 5G4, Canada}

\date{\today}

\begin{abstract}
We use cosmological perturbation theory to study the backreaction effects of a self-consistent and well-defined cosmological averaging on the dynamics and the evolution of the Universe. Working with a perturbed Friedman-Lema\^itre-Robertson-Walker Einstein-de Sitter cosmological solution in a comoving volume-preserving gauge, we compute the expressions for the expansion scalar and deceleration parameter to second order, which we use to characterize the backreaction. We find that the fractional shift in the Hubble parameter with respect to the input background cosmological model is $\overline{\Delta}\approx 10^{-5}$, which leads to $\Omega_{\rm eff}$ of the order of a few times $10^{-5}$. In addition, we find that an appropriate measure of the fractional shift in the deceleration parameter $\overline{\mathcal{Q}}$ is very large.
\end{abstract}

\maketitle

\section{Introduction}
\noindent It is currently of great interest to determine the quantitative size of backreaction effects, obtained by averaging the Einstein field gravitational field equations on cosmological scales, on the dynamics and the evolution of the Universe and consequently on cosmological observations (see \cite{BOOK,Russ:1997,Buchert:2000,Buchert:2001,Rasanen:2003,Rasanen06,Li:2007,Paranjape07,Behrend:2008,Paranjape:2008,Kolb09,Brown:2009a,Clarkson:2009,Umeh:2010pr,Clarkson:2011uk,Buchert:2011sx,Kolb:2011zz,Wiegand:2011je} for an inexhaustive sample of studies). The recent literature has begun to focus increasing attention on multi-scale averaging in cosmology (as in \cite{Wiltshire07,Wiltshire07-2,Buchert08,Wiegand:2010uh,Bose:2010gc,Bose:2013uia}) in which one or more characteristic length scales -- such as the typical scale of cosmological voids and clusters -- are used to construct a more detailed model of an averaged cosmology than is provided by a simplistic single-scale averaging procedure. However, this focus overlooks an extremely important aspect of relativistic and cosmological averaging.

In order for the results of cosmological averaging to make physical sense, it is crucial to possess a rigorous (fully covariant) definition of the spacetime average of a tensor on a differential manifold \cite{Zalaletdinov:1997,Coley:2005,Coley:2006,Coley:2007,VanDenHoogen:2007en}. In particular, it is necessary to select a coordinate system and ``gauge'' in which the averaging procedure is well-defined (as discussed in different contexts in \cite{Larena:2009,Brown:2009b,Umeh:2010pr}).

Cosmological perturbation theory provides a well-motivated paradigm in which to consider the potential quantitive effects of cosmological averaging, with parameters that can be fixed in the early, highly-uniform, universe and which have a clear physical interpretation. While any appreciable effect would suggest that cosmological perturbation theory is invalid, at least in the late universe, it does provide at least a useful toy model, so long as care is be taken when interpreting the results. (Alternative approaches employing inhomogeneous models include those in \cite{Rasanen:2004,Paranjape:2006,Marra07-2,Rasanen:2008,Vanderveld08,Bolejko:2008zv,Bolejko:2010wc,Clifton:2012qh,Boehm:2013qqa}.)

Relativistic perturbation theories are gauge-dependent. The cosmological backreaction from perturbations is therefore clearly gauge-dependent. In addition it obviously depends on the choice of averaged Hubble rate, which in a multi-fluid system is non-unique. Studies of cosmological backreaction have typically been in synchronous gauge (as in \cite{Russ:1997,Buchert:2001,Li:2007}) or conformal Newtonian gauge (as in \cite{Kasai:2006,Tanaka:2007,Behrend:2008,Brown:2009a,Clarkson:2009,Brown:2009cy,Umeh:2010pr}). Uniform curvature gauge was briefly considered in \cite{Brown:2009a}. Recently we argued that a spacetime average is only well-defined in perturbation theory when undertaken in so-called volume-preserving coordinates (VPC), or in the closely related comoving volume-preserving gauges (VPGs) \cite{Brown:2012fx} (see also \cite{Herman:2012}). In a VPC the volume of a domain is preserved as the system evolves in time.

In \cite{Brown:2012fx} we began an investigation of VPCs in cosmological averaging, focusing our attention on uniform curvature gauge, which can be interpreted as a comoving VPG. We constructed two choices of averaged Hubble rate, one naturally defined from a fluid's expansion scalar (the ``projected fluid frame'') and one from the expansion of a 3-volume (the ``gravitational frame'') and compared analytical and numerical results in the VPG with those in Newtonian gauge. A suitable measure of the impact of the ``backreaction'' is to consider the difference between the input and averaged Hubble rates. (Naturally this can be extended to other relevant observational quantities \cite{BOOK}.) After arguing that the projected fluid frame provides a more physical definition of an averaged Hubble rate, we found that the effective energy density of the backreaction, evaluated at the current epoch for an Einstein-de Sitter universe in a comoving VPG, is $\Omega_{\rm eff}\approx 5\times 10^{-4}$, which is slightly larger than but in broad agreement with previous results in conformal Newtonian gauge. However, we did not consider a measure of the acceleration rate in the comoving VPG and neither did we consider a genuine VPC, with a constant four-dimensional volume. This motivates the present study.

In this paper we extend the previous analysis. We work to second-order in perturbation theory, assuming as a toy model a pressureless Einstein-de Sitter universe. We stress that this is presented as a convenient toy model rather than necessarily an accurate model -- we are less interested in the physical viability of the models, but rather in the comparative quantitative effects of the backreaction found in consistent VPCs. In \S\ref{Sec:Dynamics} we briefly review the dynamics of a second-order Einstein-de Sitter universe in Poisson gauge (closely following \cite{Bartolo:2005kv,Matarrese:1997ay}), in which analytical solutions for the perturbations at linear- and second-order can be found, relating the Newtonian potentials at any time to the spectrum of the Newtonian potential at the present epoch.

We then discuss VPCs in cosmology in \S\ref{Sec:VPC}, in which consistent spacetime averages of cosmological perturbations can be undertaken, and present in detail the construction of a four-dimensional VPC at background, linear and second-order in perturbations. We consider an extension of the comoving VPC, or VPG, employing one of the scalar gauge freedoms to ensure that the determinant of the 4-dimensional metric is constant and then applying an appropriate coordinate transformation; this generalises the considerations in \cite{Brown:2012fx} in which we employed uniform curvature gauge to fix the determinant of the induced 3-dimensional metric to a function of the scale factor alone. The linear and second-order Newtonian potentials can then be straightforwardly transformed into this VPC.

We discuss the cosmological averaging procedure in a VPC. The backreaction is defined from the expansion scalar and deceleration parameter, which we derive to second-order. The fractional shift in the Hubble rate $\Delta$ (\ref{shiftH}) characterises the averaged Hubble rate (and could be used to estimate the effective energy density of the backreaction) while the fractional shift in the deceleration parameter $\overline{\mathcal{Q}}$ (\ref{shiftQ}) characterises the averaged deceleraton parameter. In pure FLRW evolution both of these quantities naturally vanish. Finally, we transform to Fourier space and evaluate the spatial averages employing the linear gravitational potential evaluated by a Boltzmann code developed from COSMICS and CMBFast \cite{Bertschinger:1995er,Seljak:1996}. The numerical results are presented in Figure \ref{Fig:Integrals}. We  find that the fractional shift in the Hubble parameter with respect to the input FLRW model is $\overline{\Delta}\approx 10^{-5}$  which leads to $\Omega_{\rm eff}$ of a few times $10^{-5}$. This is consistent with previous results. The averaged deceleration parameter evaluated in the VPC is found to be $\overline{\mathcal{Q}}\approx  0.44$. The size of the fluctuation to the averaged acceleration rate is very large; it is unclear to what extent this is a gauge-dependent effect and whether it is physical or not, whether it indicates a breakdown of the formalism, or whether it indicates a problem with the choice of the second scalar gauge condition. However, it can certainly be stated that averaging the deceleration parameter in a coordinate system consistent with an averaging procedure could potentially lead to a large effect, with consequences for cosmological observations or for the current understanding of perturbation theory.

\section{Dynamics}
\label{Sec:Dynamics}
\noindent In this section we briefly review the dynamics of dust-dominated Friedman-Lema\^itre-Robertson-Walker (FLRW) universes perturbed to second order. Following the notation of \cite{Malik:2008}, in an arbitrary gauge the perturbed FLRW line element is
\be
\label{FLRW}
ds^2=a^2\left(-(1+2\phi_1)d\eta^2+2B_i\eta dx^i+\left((1-2\psi_1)\delta_{ij}+2C_{ij}\right)dx^idx^j\right) .
\ee
Here
\be
B_i=\pd_iB-S_i,\quad C_{ij}=2\pd_{(i}F_{1Nj)}+\pd_{(i}F_{2Nj)}+\frac{1}{2}h_{1Nij}^{(T)}+\frac{1}{4}h_{2Nij}^{(T)}, \quad \pd^iF_i=\pd^ih_{ij}^{(T)}=\delta^{ij}h_{ij}^{(T)}=0 .
\ee
Solutions for the perturbations are relatively straightforward in Poisson gauge, in which the line element reduces to
\be
\label{FLRWPoisson}
ds^2=a^2(\eta)\left(-(1+2\phi_{1N}+\phi_{2N})d\eta^2+\left((1-2\psi_{1N}-\psi_{2N})\delta_{ij}+2C_{ij}\right)dx^idx^j\right) .
\ee
Fluid densities and pressures are expanded as
\be
\rho(\eta,x^i)=\overline{\rho}(\eta)\left(1+\delta_{1N}(\eta,x^i)+\frac{1}{2}\delta_{2N}(\eta,x^i)\right), \quad p(\eta,x^i)=\overline{p}(\eta)+\delta p_{1N}(\eta,x^i)+\frac{1}{2}\delta p_{2N}(\eta,x^i)
\ee
with a 4-velocity
\be
u^\mu=\frac{1}{a}\left(1-\phi_{1N}-\frac{1}{2}\phi_{2N}+\frac{3}{2}\phi_{1N}^2+\frac{1}{2}v^a_{1N}v_a^{1N},v^i_{1N}+\frac{1}{2}v^i_{2N}\right)
\ee
Here $v^i=\pd^iv+v^i_{(V)}$ with $\pd^av_a^{(V)}=0$.

We assume a pressureless Einstein-de Sitter universe at all times. We also neglect throughout vector and tensor modes. At linear order, vector modes decay with the scale factor and therefore rapidly become negligible, while tensor modes propagate as gravitational radiation and are suppressed by a factor of the scalar/tensor ratio, constrained to $r\lesssim 1/9$ \cite{Ade:2013uln}. At second order, in principle the scalar, vector and tensor modes are of equivalent magnitude. However, as we will later see, in the expressions we evaluate the dominant term will arise from the square of linear scalar modes, of order $\mathcal{O}(\epsilon^2)$, while the second-order vector and tensor modes will contribute quadratically, of order $\mathcal{O}(\epsilon^4)$ or in combination with linear scalar modes, of order $\mathcal{O}(\epsilon^3)$. For the purposes of averaging at second-order, vectors and tensors are therefore always subdominant to scalars.

Neglecting vector and tensor modes, the expansion scalar is
\bea
\Theta_N=\nabla^\mu u_\mu&=&\frac{1}{a}\Big(3\mH+\pd^a\pd_av_{1N}-3\psi_{1N}'-3\mH\phi_{1N}
+\frac{1}{2}\Big[\pd^a\pd_av_{2N}-3\psi_{2N}'-3\mH\phi_{2N}+9\mH\phi_{1N}^2
\nonumber\\&&\qquad
+\pd^av_{1N}\left(2\pd_av_{1N}'+3\mH\pd_av_{1N}+2\left(\pd_a\phi_{1N}-3\pd_a\psi_{1N}\right)\right)+6\left(\phi_{1N}-2\psi_{1N}\right)\psi_{1N}'\Big]\Big) .
\eea
The acceleration rate is governed by the projected time derivative of this:
\be
\label{Raychaudhuri}
\dot{\Theta}=u^\mu\nabla_\mu\Theta=u^\mu\pd_\mu\Theta
\ee
which can be recovered from the Raychaudhuri equation or directly from the expansion scalar.

In an Einstein-de Sitter universe the background solution is
\be
\overline{\rho}(\eta)=\frac{\overline{\rho}_0}{a(\eta)^3}, \quad a(\eta)=\left(\frac{\eta}{\eta_0}\right)^2, \quad \mathcal{H}=\frac{a'}{a}=\frac{2}{\eta} .
\ee
The linear system can be solved by considering the Einstein equations in Newtonian gauge \cite{Malik:2008},
\be
\begin{array}{c}
\vspace{0.5em}
3\mH\left(\psi'_{1N}+\mH\phi_{1N}\right)-\pd^a\pd_a\psi_{1N}=-4\pi Ga^2\bkr\delta_{1N}, \\
\psi'_{1N}+\mH\phi_{1N}=-4\pi Ga^2\bkr v_{1N}, \quad
\psi_{1N}=\phi_{1N}, \quad
\psi''_{1N}+3\mH\psi'_{1N}=0 .
\end{array}
\ee
Setting $\phi_{1N}=\psi_{1N}$ in the evolution equation gives
\be
\phi_{1N}=A+B\left(\frac{\eta_m}{\eta}\right)^5
\ee
This solution is initialised at a time $\eta_m$ deep in matter domination. Since we are considering the recent universe where $\eta\gg\eta_m$ (or equivalently, $a(\eta_m)\ll 1$) we can neglect the decaying mode, implying that
\be
\phi_{1N}'=0.
\ee
The Hamiltonian and momentum constraints now rapidly provide the density contrast and velocity,
\be
\label{LinearSolutions}
\delta_{1N}=-2\phi_{1N}+\frac{1}{6}\eta^2\pd^a\pd_a\phi_{1N}, \quad v_{1N}=-\frac{1}{3}\eta\phi_{1N} .
\ee

Let us turn now to the more complicated issue of the second-order scalar perturbations. Our discussion follows that of \cite{Bartolo:2005kv}, although we work in a pure EdS universe and our results are equivalent to those in \cite{Matarrese:1997ay}. The components of the $G^i_j-8\pi GT^i_j$ equation in a pure EdS universe with $\phi_{1N}'=0$ are \cite{Nakamura:2004rm,Christopherson:2011ra,Christopherson:2013} 
\bea
\label{HamiltonianConstraint}
0-0&:&3\mH\left(\psi_{2N}'+\mH\phi_{2N}\right)-\pd^a\pd_a\psi_{2N}-3\pd^a\phi_{1N}\pd_a\phi_{1N}-8\psi_{1N}\pd^a\pd_a\psi_{1N}-12\mH^2\phi_{1N}^2
\nonumber\\&&\quad
=-4\pi Ga^2\bkr\left(\delta_{2N}+2\pd^av_{1N}\pd_av_{1N}\right), \\
\label{MomentumConstraint}
0-i&:&\pd_i\psi_{2N}'+\mH\pd_i\phi_{2N}-2\mH\phi_{1N}\pd_i\phi_{1N}
=-4\pi Ga^2\bkr\left(\pd_iv_{2N}+2\pd_iv_{1N}\delta_{1N}\right), \\
\label{TraceEFE}
i-j\;\mathrm{trace}&:&\psi''_{2N}+\mH\left(2\psi'_{2N}+\phi'_{2N}\right)+\frac{1}{3}\pd^a\pd_a\left(\phi_{2N}-\psi_{2N}\right)
\\ &&\quad
-\frac{7}{3}\pd^a\phi_{1N}\pd_a\phi_{1N}-\frac{8}{3}\phi_{1N}\pd^a\pd_a\phi_{1N}=4\pi Ga^2\bkr\left(\frac{2}{3}\pd^av_{1N}\pd_av_{1N}\right), \nonumber \\
\label{TracelessEFE}
i-j\;\mathrm{traceless}&:&\pd^4\left(\psi_{2N}-\phi_{2N}\right)=12\pi Ga^2\bkr\left(2\pd_a\pd_b\left(\pd^av_{1N}\pd^bv_{1N}\right)-\pd^a\pd_a\left(\pd^bv_{1N}\pd_bv_{1N}\right)\right)
\\&&\quad\quad
-14\pd^a\pd^b\phi_{1N}\pd_a\pd_b\phi_{1N}-24\pd^a\phi_{1N}\pd_a\pd^b\pd_b\phi_{1N}-2\left(\pd^a\pd_a\phi_{1N}\right)^2-8\phi_{1N}\pd^4\phi_{1N}\nonumber
\eea
These equations can provide us with a commutator for $\psi_{2N}$ and $\phi_{2N}$ and an evolution 
equation for $\psi_{2N}$ in the same manner as the equivalent linear equations, which we derive closely following 
the method in \cite{Bartolo:2005kv}.\footnote{It should be noted that this trace corrects that presented in \cite{Bartolo:2005kv} and its references, which has an inaccurate prefactor on the $\pd^a\phi_{1N}\pd_a\phi_{1N}$ term. This appears to be a typographical error, as their final result is accurate. It should also be noted that it is possible that the decaying mode in $\phi_{1N}$ could influence the second-order perturbation theory. While the impact on the second-order potentials is expected to be negligible for dust, this should be taken into account in more general situations. Likewise, while we have neglected $\delta p_1$ due to the extremely low adiabatic speed of sound in matter domination, we have also neglected $\delta p_2$ which may contain additional contributions. The authors are grateful to Adam Christopherson for discussion of these issues.}

Defining the auxiliary functions
\be
\label{defPij}
P^i_j=2\partial^i\phi_{1N}\partial_j\phi_{1N}+8\pi Ga^2\bkr\pd^iv_{1N}\pd_jv_{1N}, \quad P=P^i_i
\ee
and the related definitions
\be
N=\pd^{-2}\left(\pd_i\pd^jP^i_j\right), \quad
\mathcal{Q}=\pd^{-2}\left(3N-P\right)
\ee
allows us to write the commutator as
\be
\psi_{2N}-\phi_{2N}=\mathcal{Q}-4\phi_{1N}^2 ,
\ee
as can be verified by expanding this equation out and comparing with equation (\ref{TracelessEFE}) with $\bkp=\delta p=0$.

Using the form of the solutions at linear order it is easy to see that
\be
\begin{array}{c}
\vspace{0.5em}
P^i_j=\dfrac{10}{3}\pd^i\phi_{1N}\pd_i\phi_{1N}, \quad N=\dfrac{10}{3}\pd^{-2}\left(\pd_i\pd^j\left(\pd^i\phi_{1N}\pd_j\phi_{1N}\right)\right), \\
\vspace{0.5em}
\mathcal{Q}=10\pd^{-2}\pd^{-2}\left(\pd_i\pd^j\left(\pd^i\phi_{1N}\pd_j\phi_{1N}\right)\right)-\dfrac{10}{3}\pd^{-2}\left(\pd^i\phi_{1N}\pd_i\phi_{1N}\right), \\
P_{ij}'=0, \quad \mathcal{Q}'=0 .
\end{array}
\ee
In particular, this implies that
\be
\psi_{2N}'=\phi_{2N}' .
\ee
(Note that this holds only because we are working in an EdS universe; in a $\Lambda$CDM universe this is no longer true and the time derivatives of the potentials will increasingly differ from one-another.)

The evolution equation from the trace can be rewritten as
\bea
\psi_{2N}''+3\mH\psi_{2N}'&=&\mH\left(\psi_{2N}'-\phi_{2N}'\right)+\frac{1}{3}\pd^a\pd_a\left(\psi_{2N}-\phi_{2N}\right)
\\ && \quad
+\frac{7}{3}\pd^a\phi_{1N}\pd_a\phi_{1N}+\frac{8}{3}\phi_{1N}\pd^a\pd_a\phi_{1N}+\frac{8\pi Ga^2}{3}\bkr\pd^av_{1N}\pd_av_{1N} .
\eea
Inserting the commutator this reduces to
\be
\psi_{2N}''+3\mH\psi_{2N}'=N-\pd^a\phi_{1N}\pd_a\phi_{1N}=S .
\ee
Since the source $S$ is time-independent this is trivially solved to give
\be
\psi_{2N}=A+B\left(\frac{\eta_m}{\eta}\right)^5+\frac{1}{14}\eta^2\left(N-\pd^a\phi_{1N}\pd_a\phi_{1N}\right) .
\ee

Boundary conditions can be set using the curvature perturbation on uniform density hypersurfaces, which is conserved on large-scales \cite{Malik:2008}.\footnote{This is, strictly speaking, not the case. In a multi-fluid system relative entropy perturbations between species induce a non-adiabatic pressure $\delta p_{\rm rel}$ which acts as a source for $\zeta_2$ \cite{GarciaBellido:1995qq,Wands:2000dp,Malik:2008,Christopherson:2011ra,Nalson:2011gc}. This non-adiabatic pressure grows until the epoch of matter/radiation equality, after which time it begins to decay again \cite{Brown:2011dn}. As a result, this initial condition can only be used to provide an order-of-magnitude estimate.} The curvature perturbation produced after inflation produces \cite{Bartolo:2005kv}
\be
\zeta_2(\eta_m)\approx \frac{50}{9}\phi_{1N}^2 .
\ee
Employing the second-order Hamiltonian constraint and the relationship between $\zeta_2$ and the Newtonian gauge variables \cite{Bartolo:2005kv,Malik:2008} then gives
\be
\phi_{2N}(\eta_m)=2\phi_{1N}^2+2\pd^{-2}\left(\pd^i\phi_{1N}\pd_i\phi_{1N}\right)-6\pd^{-4}\left(\pd_i\pd^j\left(\pd^i\phi_{1N}\pd_j\phi_{1N}\right)\right) .
\ee
Using the commutator to find $\psi_{2N}(\eta_m)$ and dropping the decaying mode allows us to fix the constant $A$, ultimately yielding
\bea
\psi_{2N}(\eta)\approx -2\phi_{1N}^2+\frac{6}{5}\pd^{-2}N-\frac{4}{3}\pd^{-2}\left(\pd^i\phi_{1N}\pd_i\phi_{1N}\right)+\frac{1}{14}\eta^2\left(N-\pd^i\phi_{1N}\pd_i\phi_{1N}\right), \\
\phi_{2N}(\eta)\approx 2\phi_{1N}^2-\frac{9}{5}\pd^{-2}N+2\pd^{-2}\left(\pd^i\phi_{1N}\pd_i\phi_{1N}\right)+\frac{1}{14}\eta^2\left(N-\pd^i\phi_{1N}\pd_i\phi_{1N}\right) .
\eea
These agree with the results of \cite{Matarrese:1997ay,Bartolo:2005kv} reduced to pure EdS and $\eta\gg\eta_m$. We also require the Laplacian of the velocity in Poisson gauge, which can be found from the momentum constraint (\ref{MomentumConstraint}) to be
\bea
\label{LaplacianVelocityNewtonian}
\pd^a\pd_av_{2N}&=&-\frac{1}{3}\eta\left(6\phi_{1N}\pd^a\pd_a\phi_{1N}+8\pd^a\phi_{1N}\pd_a\phi_{1N}-\frac{9}{5}N+\frac{1}{7}\eta^2\pd^a\pd_a(N-\pd^b\phi_{1N}\pd_b\phi_{1N})
\right. \nonumber \\ && \qquad\qquad \left.
-\frac{1}{3}\eta^2\pd^a\pd_a\phi_{1N}\pd^b\pd_b\phi_{1N}-\frac{1}{3}\eta^2\pd^a\phi_{1N}\pd_a\pd^b\pd_b\phi_{1N}\right)
 .
\eea
The expansion scalar can now be written to second order in Poisson gauge in a pure EdS universe in terms of the linear potential $\phi_{1N}$,
\be
\Theta_N=\frac{\eta_0^2}{\eta^3}\left(6(1-\phi_{1N})-\frac{1}{3}\eta^2\pd^a\pd_a\phi_{1N}+\frac{1}{2}\left(\eta\pd^a\pd_av_{2N}-3\eta\psi_{2N}'-6\phi_{2N}\right)+9\phi_{1N}^2+\frac{10}{9}\eta^2\pd^a\phi_{1N}\pd_a\phi_{1N}\right)
\ee
We calculate the acceleration rate directly from (\ref{Raychaudhuri}), giving
\bea
\dot{\Theta}_N&=&\frac{\eta_0^4}{\eta^6}\Big(
-18
+36\phi_{1N}
+\frac{1}{3}\eta^2\pd^a\pd_a\phi_{1N}
-\eta\pd^a\pd_av_{2N}
-\frac{3}{2}\eta^2\psi_{2N}''
+3\eta\left(\psi_{2N}'-\phi_{2N}'\right)
+18\phi_{2N}
\nonumber\\&&\qquad
-\frac{1}{3}\eta^2\phi_{1N}\pd^a\pd_a\phi_{1N}
-72\phi_{1N}^2
-\frac{1}{9}\eta^2\pd^a\phi_1\pd_a\phi_{1N}
+\frac{1}{9}\eta^4\pd_a\pd^b\pd_b\phi_{1N}\pd^a\phi_{1N}
\Big)
\eea

\section{Volume-Preserving Coordinate Systems in Cosmology}
\label{Sec:VPC}

\noindent\textbf{Background}\newline
\noindent Let us consider first the background cosmology, which will determine the coordinate transformation necessary to ensure an appropriate volume element and a suitable notation. By definition a volume-preserving coordinate system is one in which $\sqrt{-g}=1$. The determinant of the metric of the flat FLRW background spacetime (\ref{FLRW}) is $g=-a^8(t)$. This can be brought into a volume-preserving form through the coordinate transformation
\be
a^2d\eta^2=\frac{d\sigma^2}{a^6}\Rightarrow \sigma=\int_0^\eta a^4(\eta')d\eta' .
\ee
For a universe filled with dust we therefore have that
\be
\sigma=\frac{\eta^9}{9\eta_0^8}, \quad
\label{VPCTimeScaleFactor}
a(\sigma)=\left(\frac{\sigma}{\sigma_0}\right)^{2/9}, \quad \mHV=\frac{\dot{a}}{a}=\frac{2}{9\sigma} .
\ee
The physical Hubble rate is $H=a^3\mHV$, and an overdot denotes differentiation with respect to $\sigma$.

\vspace{1.5em}\noindent\textbf{Linear Order}\newline
\noindent We can now consider a volume-preserving coordinate system at linear order, with the procedure we follow standing as a template for the significantly more complicated non-linear case. The material in this section was first presented in \cite{Herman:2012}. The determinant of a linearly perturbed FLRW metric in an unspecified gauge is
\be
\sqrt{-g}=a^4\left(1+\phi_1-3\psi_1+\partial^i\partial_iE_1\right)
\ee
where we have not neglected vector and tensor components. A comoving VPC with $\sqrt{-g}=a^4(\eta)$ is therefore defined by choosing a gauge in which
\be
\label{GaugeCondition1}
\phi_{1V}-3\psi_{1V}+\partial^i\partial_iE_{1V}=0.
\ee
This determines one of the two scalar gauge freedoms. The other can be specified by enforcing
\be
\psi_{1V}=0 \Rightarrow \phi_{1V}=-\pd^i\pd_iE_{1V},
\ee
which ensures the gauge transformation is cleanly defined. It is important to note that there is therefore not a single VPC in cosmology but rather a family of VPCs dependent on the choice of this second remaining freedom. The gauge transform is generated by the 4-vector $\xi^\mu=(\alpha,\partial^i\beta)$, and choosing a gauge in which the spatial curvature vanishes implies
\be
\alpha_1=\frac{\psi_1}{\mathcal{H}} .
\ee
The metric determinant then transforms as
\be
\phi_{1V}+\partial^i\partial_iE_{1V}=0=\phi_1-3\psi_1+\partial^i\partial_iE_1+\alpha'_1+4\mathcal{H}\alpha_1+\partial^i\partial_i\beta_1 .
\ee
This can be solved for the spatial component of the gauge transformation,
\be
\partial^i\partial_i\beta_1=-\left(\phi_1+\psi_1+\partial^i\partial_iE_1+\left(\frac{\psi_1}{\mathcal{H}}\right)'\right) .
\ee
Applying this gauge transformation to the metric and fluid quantities therefore gives the perturbations in the comoving VPC,
\bea
\phi_{1V}=\phi_1+\psi_1+\left(\frac{\psi_1}{\mathcal{H}}\right)', \quad
\partial^i\partial_iB_{1V}=\partial^i\partial_iB_1-\frac{\partial^i\partial_i\psi_1}{\mathcal{H}}-\left(\phi_1+\psi_1+\partial^i\partial_iE_1+\left(\frac{\psi_1}{\mathcal{H}}\right)'\right)', \\
\delta_{1V}=\delta_1-3(1+w)\psi_1, \quad
\partial^i\partial_iv_{1V}=\partial^i\partial_iv_1+\left(\phi_1+\psi_1+\partial^i\partial_iE_1+\left(\frac{\psi_1}{\mathcal{H}}\right)'\right)'.
\eea
Specialising to the transformation of Newtonian-gauge dust perturbations into the comoving VPC gives
\be
\phi_{1V}=\frac{5}{2}\phi_{1N}, \quad
\pd^i\pd_iB_{1V}=-\frac{1}{2}\eta\pd^i\pd_i\phi_{1N}, \quad
\delta_{1V}=-5\phi_{1N}+\frac{1}{6}\eta^2\pd^i\pd_i\phi_{1N}, \quad
\pd^i\pd_iv_{1V}=-\frac{1}{3}\eta\pd^i\pd_i\phi_{1N} .
\ee
The metric can be brought into a true volume-preserving form with $d\sigma=a^4d\eta$, leaving the line element in a linear cosmological VPC
\be
ds^2=-\frac{1}{a^6}(1+2\phi_{1V})d\sigma^2+\frac{2}{a^2}\partial_iB_{1V}d\sigma dx^i+a^2\left(\delta_{ij}+2\partial_i\partial_jE_{1V}\right)dx^idx^j
\ee
with $\phi_{1V}=-\pd^a\pd_aE_{1V}$. Writing quantities in terms of the Newtonian-gauge potential, the expansion scalar and acceleration rate to first order are
\be
\Theta_V=\frac{\eta_0^2}{\eta^3}\left(6-15\phi_{1N}-\frac{1}{3}\eta^2\pd^a\pd_a\phi_{1N}\right), \quad
\dot{\Theta}_V=\frac{\eta_0^4}{\eta^6}\left(-18+90\phi_{1N}+\frac{1}{3}\eta^2\pd^a\pd_a\phi_{1N}\right) .
\ee

\vspace{1.5em}\noindent\textbf{Second Order}\newline
\noindent The linear VPC presented in the above section can be extended to second-order in perturbations in a straightforward manner. To second order in perturbations
\be
\sqrt{-g}=a^4(\eta)\left(1+\phi+C+\phi C+\frac{1}{2}\left(C^2+B^aB_a-\phi^2\right)-C^{ij}C_{ij}\right) .
\ee
Here $C=C^i_i=-3\psi+\partial^i\partial_iE$. As claimed earlier, vector and tensor modes enter only in products, and therefore only the linear vector and tensor perturbations are significant. Since for linear modes $F_i\propto a^{-1}$, while $\mathcal{O}(h_{ij})\sim r\mathcal{O}(\psi)$ and $r\lesssim 1/9$, the impact of these modes on the determinant can be neglected. We assume that the linear sector is already in the comoving VPC defined above, with time coordinate $\eta$. In this section, perturbations without a subscript contain both linear and second-order components; for instance, $C=C_1+(1/2)C_2$.

Finding a VPC to second order becomes significantly more straightforward if we choose to transform from a system in second-order perturbed Poisson gauge. If we later had solutions in, for instance, synchronous gauge it is straightforward to first transform them to Newtonian gauge and then the VPC. The task also becomes far easier if we also focus from the outset on dust-dominated universes, which is sufficient for our purposes but must be generalised if one wishes to consider systems containing radiative species or some form of dark energy.

As in the linear case we fix the first gauge freedom by setting
\be
\psi_{2V}=0 .
\ee
It must be emphasised again that as at linear order, there are a number of ways in which a comoving VPC can be fixed at second-order. The choice $\psi_{2V}=0$ provides a straightforward case in which
\be
\alpha_{2}=\frac{\psi_2}{\mH}+\frac{1}{4\mathcal{H}}\left(\pd^{-2}\pd_i\pd_j\chi^{ij}-\chi^k_k\right),
\ee
with $\chi_{ij}$ defined later. Since this is the same as the $\alpha_{2}$ required to transform between Poisson and uniform curvature gauges at second order (see equation (7.24) of \cite{Malik:2008}), this transformation produces gauges that are fully fixed in their time evolution. However, it is possible an alternative choice would produce a gauge better-suited to the problem. We further discuss this issue later.

The effects of averaging at second-order are known to be gauge-dependent. For example, for an appropriate choice of averaged Hubble rate, it is possible to choose a gauge in which a measure of the backreaction on the Hubble rate vanishes (as in the ``gravitational frame'' in uniform curvature gauge in \cite{Finelli:2011cw,Brown:2012fx}), or alternatively to choose gauges in which logarithmic divergences arising from superhorizon modes are also absent (as in the ``projected fluid frame'' in uniform curvature gauge in \cite{Brown:2012fx}). While the formalism we will employ in this paper is different from that employed in these earlier studies, such logarithmic divergences are endemic in studies of second-order perturbations and should be expected to naturally arise here. In principle, we could choose the second gauge condition to explicitly eliminate logarithmic divergences in the averaged expansion or acceleration rate $q=-\dot{H}/H^2-1$, although it may not prove possible to simultaneously eliminate them from both quantities.

In any event, the logarithmic divergences are clearly gauge effects given they arise in some formulations (for instance, \cite{Clarkson:2009,Umeh:2010pr}) and not in others (as in \cite{Brown:2012fx}), and should not be expected to influence the physical results. On a practical level, we can handle them by imposing an infra-red cut-off scale at or above the Hubble scale; since the divergence is typically logarithmic the impact of changing the cut-off even by an order of magnitude is not very significant. We will return to this issue later.

In this paper, to keep things relatively simple and not overly complicate the analysis, we shall consider the simpler case with $\psi_{2V}=0$. Neglecting the vectors and tensors, expanding out the determinant and enforcing the VPG condition at linear order gives
\be
\sqrt{-g}=a^4(\eta)\left(1+\frac{1}{2}\phi_{2V}+\frac{1}{2}\pd^i\pd_iE_{2V}+\frac{1}{2}\pd^iB_{1V}\pd_iB_{1V}-\pd^i\pd^jE_{1V}\pd_i\pd_jE_{1V}-\pd^i\pd_iE_{1V}\pd^j\pd_jE_{1V}\right) .
\ee
The spatial part of the transformation $\beta_{2}$ enters only through the second order anisotropic stress scalar transformation, equation \ref{transE2}. We can also see that for a dust system in Poisson gauge, $\beta_{1}'=0$. Inserting the gauge transformations for $\phi_2$ (\ref{transPhi2}), $E_2$ (\ref{transE2}) and substituting the dust solutions for the linear perturbations (\ref{LinearSolutions}) allows us to solve for $\beta_2$, giving
\bea
\label{VPCBeta2} 
\pd^i\pd_i\beta_2=-\Big(\phi_{2N}+\psi_{2N}+\alpha_{2}'+\pd^{-2}\pd_i\pd_j\X^{ij}-\frac{1}{2}\X_k^k-4\phi_{1N}^2+\frac{1}{4}\eta^2\pd^i\phi_{1N}\pd_i\phi_{1N} \nonumber\\
-\frac{35}{4}\pd_i(\pd^{-2}\phi_{1N})\pd^i\phi_{1N}-\frac{25}{2}\pd_i\pd_j(\pd^{-2}\phi_{1N})\pd^i\pd^j(\pd^{-2}\phi_{1N}) \Big).
\eea
The quadratic $\chi$ terms (\ref{Xdef}, \ref{XBdef}, \ref{Xvdef}) for this transformation, the perturbations at second order and other useful quantities are presented in Appendix \ref{Appendix-VPCDynamics}.

The expansion scalar and acceleration rate become to second order
\bea
\label{VPCTheta}
\Theta_V&=&\frac{\eta_0^2}{\eta^3}\left(
6-15\phi_{1N}-\frac{1}{3}\eta^2\pd^a\pd_a\phi_{1N}
+\frac{1}{2}\left[\eta\pd^a\pd_av_{2N}
-3\eta\psi_{2N}'
-3\left(3\psi_{2N}+2\phi_{2N}\right)
+\frac{99}{2}\phi_{1N}^2
+\frac{20}{9}\eta^2\pd^a\phi_{1N}\pd_a\phi_{1N}
\right.\right.\nonumber\\&&\quad\left.\left.
+\frac{9}{4}\X^k_k-\frac{9}{4}\pd^{-2}\pd_i\pd_j\X^{ij}
+\frac{1}{3}\eta^2\phi_{1N}\pd^a\pd_a\phi_{1N}
+\frac{5}{3}\eta^2\pd_k\pd^a\pd_a\phi_{1N}\pd^k\pd^{-2}\phi_{1N}
+\frac{105}{2}\pd_k\phi_{1N}\pd^k\pd^{-2}\phi_{1N}
\right]\right), \\
\dot{\Theta}_V&=&\frac{\eta_0^4}{\eta^6}\left(
-18+90\phi_{1N}+\frac{1}{3}\eta^2\pd^a\pd_a\phi_{1N}
+\frac{1}{2}\left[
\eta^2\pd^a\pd_av_{2N}'
-2\eta\pd^a\pd_av_{2N}
-3\eta^2\psi_{2N}''
+6\eta\left(\psi_{2N}'-\phi_{2N}'\right)
\right.\right.\nonumber\\&&\quad\left.\left.
+18\left(3\psi_{2N}+2\phi_{2N}\right)
-522\phi_{1N}^2
-2\eta^2\phi_{1N}\pd^a\pd_a\phi_{1N}
-\frac{2}{9}\eta^2\pd^a\phi_{1N}\pd_a\phi_{1N}
+\frac{2}{9}\eta^4\pd^a\phi_{1N}\pd_a\pd^b\pd_b\phi_{1N}
\right.\right.\nonumber\\&&\quad\left.\left.
-\frac{5}{3}\eta^2\pd_k\pd^a\pd_a\phi_{1N}\pd^k\pd^{-2}\phi_{1N}
-315\pd^k\phi_{1N}\pd_k\pd^{-2}\phi_{1N}
+\frac{27}{2}\pd^{-2}\pd_i\pd_j\X^{ij}
-\frac{27}{2}\X^k_k
\right]\right) .
\eea
These quantities can be transferred into the full VPC with the substitution $\eta=\gamma\sigma^{1/9}$.

With this coordinate transformation $\xi=(\alpha_2,\pd_i\beta_2)$ the metric is now in VPC form to second order,
\be
ds^2=-\frac{1}{a^6}\left(1+2\phi_{1V}+\phi_{2V}\right)d\sigma^2+\frac{1}{a^2}\partial_i(2B_{1V}+B_{2V})d\sigma dx^i+a^2\left(\delta_{ij}+2\partial_i\partial_jE_{1V}+\pd_i\pd_jE_{2V}\right)dx^idx^j
\ee
where
\be
\phi_{1V}=-\pd^a\pd_aE_{1V}, \quad
\phi_{2V}=-\left(\pd^a\pd_aE_{2V}+\pd^aB_{1V}\pd_aB_{1V}-2\pd^a\pd^bE_{1V}\pd_a\pd_bE_{1V}-2\pd^a\pd_aE_{1V}\pd^b\pd_bE_{1V}\right) .
\ee

\section{Spacetime Averaging in a Cosmological VPC}
\subsection{Averaging and Measures of Deviations from FLRW Form (``Backreaction'')}
\noindent A 4-domain in a 3+1 split can be defined as a 4-cylinder bounded by a range of the rescaled time coordinate $\sigma$ and a 3-domain $\mathcal{D}$, $\Sigma=(\sigma+\sigma')\times\mathcal{D}$ where $\sigma'$ is a small perturbation on the time $\sigma$ and $\mathcal{D}$ is a three-dimensional domain. The average of a scalar quantity in the VPC is then
\be
\av{A(x^\mu)}_4=\frac{1}{V_4}\av{A(x^\mu)d^4x}=\frac{1}{V_4}\iint A(\sigma+\sigma'\hspace{-0.2em},x^i)d\sigma'dx^i
\ee
where the 4-volume is $V_4=\int d^4x$ and the spatial average is implicitly taken to lie in the domain $\mathcal{D}$. Within the assumptions of the cosmological models under consideration, evolution is smooth with respect to the time coordinate $\sigma$; therefore, there is no appreciable small-scale deviation of $A(\sigma,x^i)$ with respect to timescales $\sigma'\ll\sigma$. Thus, writing
\be
\int\left(\int A(\sigma,x^i)d^3x\right)d\sigma'=\alpha\int A(\sigma,x^i)d^3x
\ee
for some $\alpha$ determined by the 4-domain, and therefore with
\be
V_4=\iint d\sigma'd^3x=\alpha V_3
\ee
we have that
\be
\av{A(x^\mu)}_4=\frac{\alpha\int A(\sigma,x^i)d^3x}{\alpha V_3}=\av{A(\sigma,x^i)}_3 .
\ee
For a perturbed FLRW model, the 4-dimensional average reduces to a spatial average.\footnote{Note that while this closely resembles an average in uniform curvature gauge, this is \emph{not} the average on the surfaces of constant $\sigma$ -- the induced metric on those surfaces, $h_{ij}=h^\mu_ih^\nu_jg_{\mu\nu}$, is not flat and the volume element is $\sqrt{h}d^3x$.}

The most natural way to characterise the expansion of a fluid is through its expansion scalar $\Theta=\nabla^\mu u_\mu$. This has been employed in the Buchert formalism, averaged across the fluid's rest-frame (e.g. \cite{Buchert:2001,Rasanen:2003,Li:2007,Umeh:2010pr}). A deviation from pure FLRW evolution can be characterised by
\be
\Delta=\frac{\av{\Theta_V}_4-\av{\Theta_0}_4}{\av{\Theta_0}_4}=\frac{\av{\delta\Theta_V}_3}{\Theta_0}
\ee
The 3-average is tractable in a similar manner to those taken in a more typical approach (e.g. \cite{Buchert:2001,Li:2007,Brown:2009a,Clarkson:2009,Umeh:2010pr,Brown:2012fx} and their references).

The other characteristic of the evolution of an FLRW model can be parameterised by
\be
q=-3\frac{\dot{\Theta}}{\Theta^2}-1
\ee
where an overdot here denotes the covariant derivative projected along the fluid 4-velocity, $\dot{\Theta}=u^\mu\nabla_\mu\Theta$. In an EdS universe, $q_0=1/2$. Unlike the expansion scalar, $q$ is dimensionless. We can characterise the average acceleration by defining
\be
\mathcal{Q}=\frac{\av{q_V}_4-\av{q_0}_4}{\av{q_0}_4}=\frac{2\av{\delta q}_3}{q_0} .
\ee
It is not our intention to construct a consistent effective FLRW cosmology based on the backreaction terms. Rather, we are interested in characterising the magnitude of deviations from the FLRW background. In this context, $\Delta$ and $\mathcal{Q}$ provide ideal test variables, which can be readily interpreted. If an interpretation in terms of some effective dynamical model is required, the simple commutation between spatial averaging and time differentiation that arises in a VPC would make this a relatively straightforward procedure.

The impact from linear tensor modes was considered in \cite{Brown:2009cy}, in which they were found to provide an effect that was unsurprisingly weak, but which acted as a fluid with $w_{\rm eff}\approx -8/9$. Although that study considered 3-averages, the effective energy density and pressure in this study were found from spatial averages which were effectively in uniform curvature gauge, with the corrections to the volume element from the gravitational waves entirely negligible. The contribution of the linear tensor modes to the second-order scalar perturbations is likewise extremely subdominant to that from linear scalar modes, and the scalar and tensors therefore do not couple to any appreciable degree. The results of \cite{Brown:2009cy} can consequently be directly employed.

\subsection{Consistent Spacetime Averages of Cosmological Perturbations}
\subsubsection{Spacetime Averages of Perturbations}
\noindent Spacetime averaging is most conveniently performed in Fourier space, in which the above forms simplify considerably and in which we possess both numerical and analytic solutions for the linear Newtonian potential $\phi_{1N}$. The transformation is defined as
\be
A(\xv)=\int A(\kv)\exp(-i\kv\cdot\xv)\dtk, \quad A(\kv)=\int A(\xv)\exp(i\kv\cdot\xv)d^3x .
\ee
Defining the 3-volume with a window function allows us to perform the integral over $x^i$ across all scales. Letting $W(\xv)$ be the window function defining the spatial domain, the spatial average and 3-volume are then
\be
\av{A(\xv)}_3=\frac{1}{V_3}\int W(\xv)A(\xv)d^3x, \quad V_3=\int W(\xv)d^3x .
\ee
The Fourier transform of the window function is $W(\kv)=\int W(\xv)\exp(i\kv\cdot\xv)d^3x$ implying $W(\kv=\mathbf{0})=\int W(\xv)d^3x=V_3$. That is, the 3-volume is the zero mode of $W(\kv)$. The 3-average is therefore
\bea
\av{A(\xv)}_3&=&\frac{1}{V_3}\iiint W^*(\kv')A(\kv)\exp(i(\kv'-\kv)\cdot\xv)\frac{d^3\kv'}{(2\pi)^3}\dtk d^3x
=\frac{1}{V_3}\iint W^*(\kv')A(\kv)\delta(\kv'-\kv)d^3\kv'\dtk
\nonumber\\
&=&\frac{1}{V_3}\int W^*(\kv)A(\kv)\dtk .
\eea
Similarly,
\be
\label{AverageConvolution}
\av{A(\xv)B(\xv)}_3=\frac{1}{V_3}\iint W^*(\kv)A(\kv')B(\kv-\kv')\frac{d^3\kv'}{(2\pi)^3}\dtk
\ee

Since our solutions at linear order only provide us with the \emph{statistical} nature of the distribution, 
to recover meaningful answers we take ensemble averages of the spatial averages. (On large scales these averages 
will tend to converge through ergodicity.) Denoting an ensemble average with an overbar, the ensemble averages of linear perturbations and products are
\be
\overline{A_1(\kv)}=0, \quad
\overline{A_1(\kv)B_1^*(\mathbf{k'})}=\frac{2\pi^2}{k^3}\mathcal{P}(k)A_1(k)B_1^*(k)(2\pi)^3\delta(\mathbf{k-k}')
\ee
with primordial power spectrum $\mathcal{P}(k)=A_\star\left(k/k_\star\right)^{n_s-1}$. An immediate well-known consequence is that the pure linear terms in $\Theta_{1V}$ and $\dot{\Theta}_{1V}$ will not contribute to the deviations from FLRW behaviour. However, they will contribute on smaller scales; when an ensemble average is not being taken it should be expected that a spatial average of the linear terms on smaller scales would contribute significantly. This will be explored in a forthcoming paper \cite{Brown:2013a}.

It is convenient to consider the separate cases of second-order averages individually.

\vspace{1.5em}\textit{Quadratic Products of Linear Perturbations}\vspace{0.3em}\newline
\noindent Consider terms of the form $\av{A(\xv)B(\xv)}_3$. The perturbations can be straightforwardly expanded in Fourier space and an ensemble average taken, to yield the familiar
\be
\label{GeneralQuadratic}
\overline{\av{A(\xv)B(\xv)}}_3=\int\mathcal{P}(k)A(k)B^*(k)\frac{dk}{k} .
\ee

\vspace{1.5em}\textit{Products of Gradients and Laplacians}\vspace{0.3em}\newline
\noindent Let $A(\xv)=\pd^a\cdots\pd_b\phi_{1N}$ and $B(\xv)=\pd^c\cdots\pd_d\phi_{1N}$, where the indices balance such that the result is a scalar, and let there be $m$ operators acting on $A(\xv)$ and $n$ acting on $B(\xv)$. Then
\be
\label{GradientsLaplacians}
\overline{\av{\pd^a\cdots\pd_b\phi_{1N}\pd^c\cdots\pd_d\phi_{1N}}}_3=\int\mathcal{P}(k)(-ik)^m(ik)^n\left|\phi_{1N}\right|^2\frac{dk}{k}
=(-1)^mi^{m+n}\int\mathcal{P}(k)k^{m+n}\left|\phi_{1N}\right|^2\frac{dk}{k} .
\ee
In particular,
\be
\overline{\av{\pd^a\phi_{1N}\pd_a\phi_{1N}}}_3=-\overline{\av{\phi_{1N}\pd^a\pd_a\phi_{1N}}}_3=\int\mathcal{P}(k)k^2\left|\phi_{1N}\right|^2\frac{dk}{k} .
\ee

\vspace{1.5em}\textit{Products containing Inverse Laplacians}\vspace{0.3em}\newline
\noindent Now let $A(\xv)$ contain an inverse Laplacian term acting on the linear perturbation. In this instance, we require the Fourier transform of $\pd^a\cdots\pd^b\pd^{-2}\pd_c\cdots\pd_d\phi_{1N}$. Consider first the case $\pd^{-2}C(\xv)$. Then we want to find the Fourier transform of $\pd^a\pd_aX(\xv)=C(\xv)$,
\be
-\int k^2X(\kv)\exp(-i\kv\cdot\xv)\dtk = \int C(\kv)\exp(-i\kv\cdot\xv)\dtk \Rightarrow X(\kv)=-\frac{1}{k^2}C(\kv);
\ee
that is, the Fourier transform of $\pd^{-2}C(\xv)$ is $-(1/k^2)C(\kv)$. Inserting $C(\xv)=\pd_a\phi_{1N}$ does not change the argument and simply inserts an additional power of $(-ik)$. Likewise, setting $A(\xv)=\pd_a\pd^{-2}C(\xv)$ in (\ref{GeneralQuadratic}) merely adds an extra power of $-ik^{-1}$. We can therefore state that
\be
\overline{\av{\pd^a\cdots\pd_b\pd^{-2p}\phi_{1N}\pd^c\cdots\pd_d\pd^{-2q}\phi_{1N}}}_3=(-1)^{m+p+q}i^{m+n}\int\mathcal{P}(k)k^{m+n-2p-2q}\left|\phi_{1N}\right|^2\frac{dk}{k} .
\ee
For instance, setting $m=1$, $n=3$, $p=q=1$ gives
\be
\overline{\av{\pd^a\pd^{-2}\phi_{1N}\pd_a\pd^b\pd_b\pd^{-2}\phi_{1N}}}_3=-\int\mathcal{P}(k)\left|\phi_{1N}\right|^2\frac{dk}{k}
\ee
as can be confirmed by direct calculation.

\vspace{1.5em}\textit{Inverse Laplacians of Products}\vspace{0.3em}\newline
\noindent These cases need a bit more care since the inverse Laplacian is now acting on a product of perturbations and the simple arguments presented above do not apply. We follow the approach of \cite{Clarkson:2009}, in which it was demonstrated (their equation (91)) that
\be
\overline{\av{\pd^{-2}\pd_i\pd_j\left(\pd^iA(\xv)\pd^jB(\xv)\right)}}_3=\frac{1}{3}\int\mathcal{P}(k)k^2A(k)B^*(k)\frac{dk}{k}
\ee
and, in particular, (their (93)) that
\be
\label{ChrisPd2}
\overline{\av{\pd^{-2}\pd_i\pd_j\left(\pd^i\phi_{1N}\pd^j\phi_{1N}\right)}}_3=\frac{1}{3}\overline{\av{\pd^a\phi_{1N}\pd_a\phi_{1N}}}_3=\frac{1}{3}\int\mathcal{P}(k)k^2\left|\phi_{1N}\right|^2\frac{dk}{k}
\ee
and (their equation (95)) that
\be
\label{ChrisPd4}
\overline{\av{\pd^{-4}\pd_i\pd_j\left(\pd^i\phi_{1N}\pd^j\phi_{1N}\right)}}_3=\frac{3}{10}\overline{\av{\pd^{-2}N}}_3=\frac{1}{3}\overline{\av{\pd^{-2}\left(\pd^a\phi_{1N}\pd_a\phi_{1N}\right)}}_3 .
\ee
These forms, along with straightforward substitutions for $A$ or $B$ of the form $A\rightarrow \pd^a\cdots\pd^b\pd^{-2}\cdots\pd^c\phi_{1N}$, are sufficient for almost all of our purposes.

The final form we require is
\be
\overline{\av{\pd^{-2}\pd_i\pd_j\left(A\pd^i\pd^jB\right)}}_3=-\frac{1}{3}\int\mathcal{P}(k)k^2\left|\phi_{1N}\right|^2\frac{dk}{k} .
\ee
This is demonstrated in Appendix \ref{Appendix-FourierSpace}.

\vspace{1.5em}\textit{Second-Order Perturbations}\vspace{0.3em}\newline
We can use the above results to evaluate the averages of $\psi_{2N}$, $\psi_{2N}'$, $\psi_{2N}''$, $\phi_{2N}$ and $\pd^a\pd_av_{2N}$, which appear in the expansion scalar and its time derivative. The Laplacian of the velocity contains the terms $\av{N}$ and $\av{\pd^a\pd_a(N-\pd^b\phi_{1N}\pd_b\phi_{1N})}_3$; the other contributions are straightforwardly recovered from the above.

The average of $N$ is also straightforward. Using (\ref{ChrisPd2}) it is easy to see that
\be
\overline{\av{N}}_3=\frac{10}{3}\overline{\av{\pd^{-2}\pd_i\pd_j\left(\pd^i\phi_{1N}\pd^j\phi_{1N}\right)}}_3=\frac{10}{9}\int\mathcal{P}(k)k^2\left|\phi_{1N}\right|^2\frac{dk}{k} .
\ee
The Laplacian of the source term, $\av{\pd^a\pd_a(N-\pd^b\phi_{1N}\pd_b\phi_{1N})}_3$, is similarly straightforward. Expanding out the derivatives and using (\ref{GradientsLaplacians}) it is easy to see that
\be
\overline{\av{\pd^a\pd_aN}}_3=\frac{10}{3}\overline{\av{\pd_i\pd_j\left(\pd^i\phi_{1N}\pd^j\phi_{1N}\right)}}_3=0
\ee
and similarly
\be
\overline{\av{\pd^a\pd_a(\pd^b\phi_{1N}\pd_b\phi_{1N})}}_3=0 .
\ee
We can then evaluate the other contributions to $\av{\pd^a\pd_av_{2N}}_3$ from (\ref{LaplacianVelocityNewtonian}) to find that
\be
\overline{\av{\pd^a\pd_av_{2N}}}_3=0
\ee
as previously seen in \cite{Clarkson:2009}.

In a similar vein, the expressions for $\psi_{2N}$ and $\phi_{2N}$ contain a term proportional to $\pd^{-2}N$. Using (\ref{ChrisPd4}) this can be written
\be
\overline{\av{\pd^{-2}N}}_3=\frac{10}{9}\overline{\av{\pd^{-2}\left(\pd^a\phi_{1N}\pd_a\phi_{1N}\right)}}_3 .
\ee
As noted in \cite{Clarkson:2009}, terms proportional to $\pd^{-2}N$ are always balanced by a term proportional to $\pd^{-2}(\pd^a\phi_{1N}\pd_a\phi_{1N})$. Therefore
\bea
\overline{\av{\psi_{2N}}}_3&=&\overline{\av{-2\phi_{1N}^2+\frac{6}{5}\pd^{-2}N-\frac{4}{3}\pd^{-2}\left(\pd^a\phi_{1N}\pd_a\phi_{1N}\right)+\frac{1}{14}\eta^2(N-\pd^a\phi_{1N}\pd_a\phi_{1N})}}_3
\nonumber\\
&=&\overline{\av{-2\phi_{1N}^2+\frac{1}{14}\eta^2(N-\pd^a\phi_{1N}\pd_a\phi_{1N})}}_3=\int\mathcal{P}(k)\left(-2+\frac{1}{126}k^2\eta^2\right)\left|\phi_{1N}\right|^2\frac{dk}{k}
\eea
and
\bea
\overline{\av{\phi_{2N}}}_3&=&\overline{\av{2\phi_{1N}^2-\frac{9}{5}\pd^{-2}N+2\pd^{-2}\left(\pd^a\phi_{1N}\pd_a\phi_{1N}\right)+\frac{1}{14}\eta^2(N-\pd^a\phi_{1N}\pd_a\phi_{1N})}}_3
\nonumber\\
&=&\overline{\av{2\phi_{1N}^2+\frac{1}{14}\eta^2(N-\pd^a\phi_{1N}\pd_a\phi_{1N})}}_3=\int\mathcal{P}(k)\left(2+\frac{1}{126}k^2\eta^2\right)\left|\phi_{1N}\right|^2\frac{dk}{k} .
\eea
Finally, we can quickly note that
\be
\overline{\av{\psi_{2N}'}}_3=\frac{1}{7}\eta\overline{\av{N-\pd^a\phi_{1N}\pd_a\phi_{1N}}}_3=\frac{1}{63}\eta\int\mathcal{P}(k)k^2\left|\phi_{1N}\right|^2\frac{dk}{k} = \eta\overline{\av{\psi_{2N}''}}_3 .
\ee

\subsubsection{The Expansion Scalar and Deceleration Parameter}
\noindent We are now in a position to evaluate the spatial averages in the averaged expansion scalar (\ref{VPCTheta}) and deceleration parameter. Background terms are invariant under the spatial averaging, while $\overline{\av{\Theta_{1V}}}_3=0$. Using (\ref{Xkk}) we can also see that
\be
\overline{\av{\X^k_k}}_3=\int\mathcal{P}(k)\left(-\frac{29}{2}-\frac{1}{2}k^2\eta^2\right)\left|\phi_{1N}\right|^2\frac{dk}{k}
\ee
while the inverse Laplacian of $\pd_i\pd_j\X^{ij}$ (\ref{invLXij}) resolves to
\be
\overline{\av{\pd^{-2}\pd_i\pd_j\X^{ij}}}_3=-\int\mathcal{P}(k)\left(\frac{29}{6}+\frac{1}{6}k^2\eta^2\right)\left|\phi_{1N}\right|^2\frac{dk}{k} .
\ee
With the results from the previous section, we can find that the fractional shift in the Hubble parameter with respect to the input FLRW model is
\be
\label{shiftH}
\overline{\Delta}=\frac{\overline{\av{\Theta_V}}_4-\overline{\av{\Theta_0}}_4}{\overline{\av{\Theta_0}}_4}
=\frac{\overline{\av{\Theta_V}}_3-\Theta_0}{\Theta_0}=\frac{1}{16}\int\mathcal{P}(k)\left(\frac{95}{27}k^2\eta^2-
25\right)\left|\phi_{1N}\right|^2\frac{dk}{k} .
\ee

The deceleration parameter evaluated in the VPC can be seen to be
\bea
q_V&=&-3\frac{\dot{\Theta}_V}{\Theta_V^2}-1
\nonumber\\
&=&\frac{1}{2}+\frac{5}{36}\eta^2\pd^a\pd_a\phi_{1N}
-\frac{1}{2}\Bigg\{
\frac{1}{12}\eta^2\pd^a\pd_av_{2N}'
+\frac{1}{3}\eta\pd^a\pd_av_{2N}
-\frac{1}{4}\eta^2\psi_{2N}''
-\frac{1}{2}\eta\left(2\psi_{2N}'+\phi_{2N}'\right)
\nonumber\\&&\quad
-\frac{5}{9}\eta^2\phi_{1N}\pd^a\pd_a\phi_{1N}
+\frac{59}{54}\eta^2\pd^a\phi_{1N}\pd_a\phi_{1N}
-\frac{7}{324}\eta^4\pd^a\pd_a\phi_{1N}\pd^b\pd_b\phi_{1N}
\nonumber\\&&\quad
+\frac{1}{54}\eta^4\pd^a\phi_{1N}\pd_a\pd^b\pd_b\phi_{1N}
+\frac{25}{36}\eta^2\pd_k\pd^a\pd_a\phi_{1N}\pd^k\pd^{-2}\phi_{1N}
\Bigg\}
\eea
which generates the fractional shift
\be
\label{shiftQ}
\overline{\mathcal{Q}}=\frac{\overline{\av{q}}_3-q_0}{q_0}=\frac{1}{54}\int\mathcal{P}(k)\left(\frac{13}{6}k^2\eta^2
-125\right)k^2\eta^2\left|\phi_{1N}\right|^2\frac{dk}{k} .
\ee
It is significant to note that unlike $\overline{\Delta}$ there is no term in the integrand proportional to 
$\left|\phi_{1N}\right|^2$! As we demonstrate in the next section, this implies that there is no large-scale logarithmic divergence in the integral and it remains safely finite -- there is no need for an unphysical infra-red cut-off. However, the integrand will instead exhibit an ultra-violet divergence, which we will control with a smoothing of the gravitational potential for scales above a smoothing scale $R_S$.

\subsubsection{Numerical Results}
\noindent On large scales, the integrand for $\overline{\Delta}$ scales as $k^{-1}$, which produces the familiar logarithmic divergence $\overline{\Delta}\sim\ln(k_{\mathrm{IR}})$ which emerges naturally in studies of nonlinearities in cosmology (as, for example, in a different context in \cite{Byrnes:2010yc}). We control this by imposing an infra-red cut-off at ten times the Hubble radius, as in \cite{Clarkson:2009}. While the choice of $k_\mathrm{IR}$ is arbitrary, the divergence is only logarithmic and there is little difference in $\overline{\Delta}$ for choices of $k_\mathrm{IR}$ within an order of magnitude of the horizon scale. There is no infra-red divergence in $\overline{\mathcal{Q}}$.

Conversely, on small scales the integrand for $\overline{\mathcal{Q}}$ scales as $(\ln k)^2/k$. This produces an ultra-violet divergence $\overline{\mathcal{Q}}\sim(\ln k_\mathrm{UV})^3$. We control this as with similar integrals in \cite{Clarkson:2009,Umeh:2010pr,Brown:2012fx} by applying a smoothing function
\be
W(kR_S)=\exp(-k^2R_S^2)
\ee
to the gravitational potential. The most physical choice of smoothing scale $R_S$ is the Silk scale which for this model is $R_S\approx 6\mathrm{Mpc}$. The ultra-violet divergence does not appear in $\overline{\Delta}$\footnote{Note that in \cite{Brown:2012fx} neither divergence appeared for quantities evaluated in the uniform curvature gauge, nor for the Newtonian gauge in the projected fluid frame.} but we present results both with the integrand smoothed, for consistency with $\overline{Q}$, and left unsmoothed.

\begin{figure}
\includegraphics[width=0.5\textwidth]{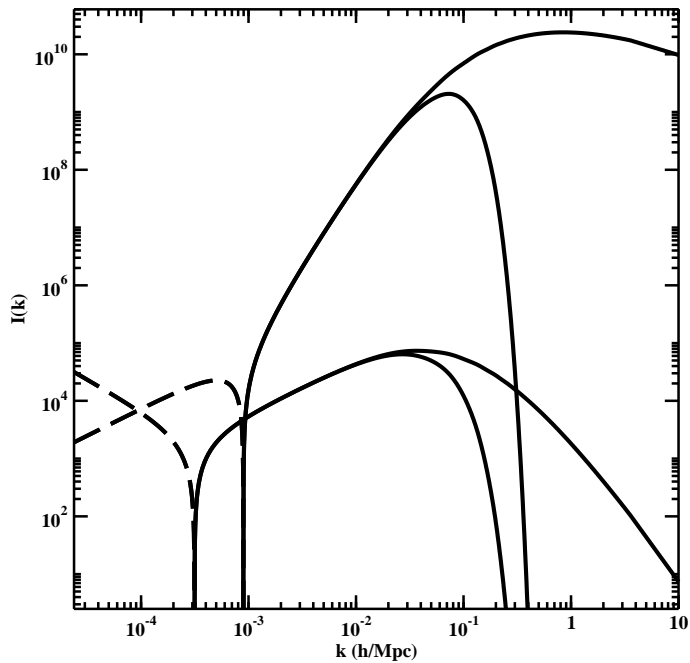}
\caption{Integrands of $\overline{\Delta}$ (lower curves) and $\overline{\mathcal{Q}}$ (upper curves) with and without smoothing at the Silk damping scale. Dashed lines are negative.}
\label{Fig:Integrands}
\end{figure}

\begin{figure}
\includegraphics[width=0.5\textwidth]{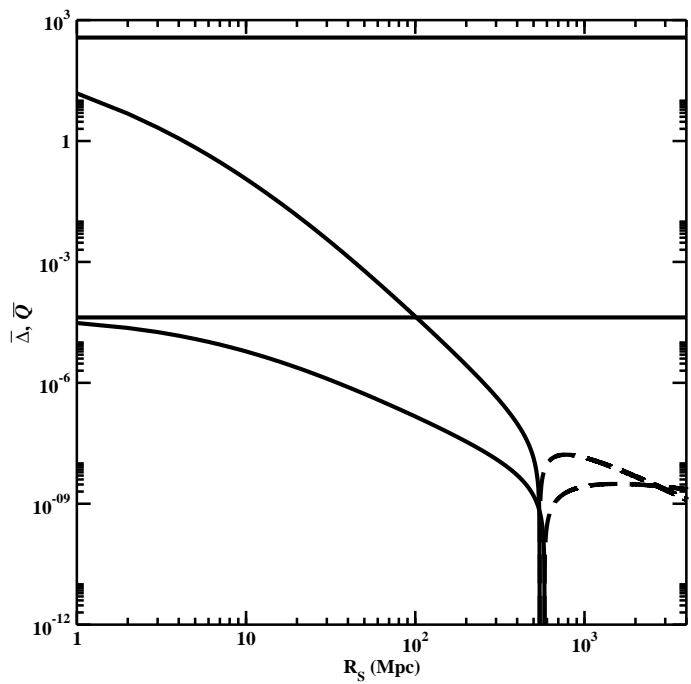}
\caption{$\overline{\Delta}$ and $\overline{\mathcal{Q}}$ for a range of smoothing scales $R_S$. Constant lines denote the case with no smoothing. Note that in this case the value of $\overline{\mathcal{Q}}$ is arbitrary; the plotted values are with $k_{\mathrm{max}}=10h/\mathrm{Mpc}$.}
\label{Fig:Integrals}
\end{figure}

Figure \ref{Fig:Integrands} shows the integrands for $\overline{\Delta}$ and $\overline{\mathcal{Q}}$ for $R_S=R_\mathrm{Silk}$. Solving the integrals numerically, we find that
\be
\overline{\Delta}=\left\{\begin{array}{rl}4.2\times 10^{-5},&\mathrm{not}\;\mathrm{smoothed} \\ 1.0\times 10^{-5},&\mathrm{smoothed}\end{array}\right.,
\quad
\overline{\mathcal{Q}}=0.44 .
\ee
For such a small value of $\overline{\Delta}$ this implies
\be
\Omega_{\rm eff}\approx\left\{\begin{array}{rl}8\times 10^{-5},&\mathrm{not}\;\mathrm{smoothed} \\ 2\times 10^{-5},&\mathrm{smoothed}\end{array}\right.
\ee
In Figure \ref{Fig:Integrals} we plot $\overline{\Delta}$ and $\overline{\mathcal{Q}}$ as functions of the smoothing scale $R_S$. To reduce the impact of the perturbations on $\av{q}_4$ to the level of $10^{-5}$ requires the perturbations to be damped on scales as large as $R_S\gtrsim 100$Mpc/h. As $R_S$ grows beyond the homogeneity scale $100-115\rm{Mpc}$ \cite{Scrimgeour:2012wt} the impacts decay rapidly to zero, as one should expect.

The direct impact on the Hubble rate is therefore of a similar order of magnitude to previous estimates. Conversely, the shift in the acceleration rate is extremely large -- of the order of 50\% -- driven by the rapid ultra-violet divergence that generates the large peak visible in the integrand. It is also interesting to note that $\mathcal{Q}>0$, implying that $q$ is moved \emph{further} from zero, and while this result may na\"ively suggest a large impact on the background acceleration it cannot immediately be connected with the observed acceleration, for which we would desire $\overline{\mathcal{Q}}\approx -1$ if we wished an accelerating averaged EdS universe. 
It is unclear to what extent this is a gauge-dependent effect, and given the magnitude of the result the study of a volume-preserving coordinate system with the second gauge choice fixed through physical arguments seems imperative. Cosmological averaging is only properly defined when working within a VPC, but we have only considered the simplest possible on here. It is now a matter of importance to investigate whether $\overline{\mathcal{Q}}\sim 0.4$ is a generic prediction, or whether a VPC can be found in which such divergences are eliminated from both $\overline{\Delta}$ and $\overline{\mathcal{Q}}$. The calculations in this paper form a vital first step in this procedure.

\section{Discussion}
\noindent In this paper we have considered the fractional shifts in the Hubble and deceleration parameters, averaged across a spatial volume. We have advocated the use of volume-preserving coordinate systems, in which consistent spacetime averages of cosmological perturbations can be undertaken. These VPCs are defined through two scalar gauge freedoms, one of which is used to set the metric determinant to unity. The other gauge freedom is arbitrary, and there is therefore an infinite family of cosmological VPCs. Choosing the second gauge condition to set the spatial curvature to zero -- a choice motivated by convenience, since it provides the simplest system to be solved for the gauge generating vector -- we then constructed a VPC valid up to second-order in perturbation theory. Cosmological perturbations in an Einstein-de Sitter universe perturbed to second-order can then be found in the VPC in a straightforward manner by transforming them from the Newtonian gauge, in which they are well-known. Spacetime averaging is then easily defined.

This approach is complementary to other research in the area, which extends the usual single-scale cosmological averaging to multiple scales, reflecting the hierarchy of scales apparent in the observations. It also has consequences for averaging in unimodular gravity, which is equivalent to general relativity but naturally formulated in volume-preserving coordinate systems \cite{Unimod}.

Deriving the expansion and deceleration parameters for dust-dominated universes perturbed to second-order, we characterised the backreaction through the fractional shifts in the Hubble parameter, $\Delta$ (\ref{shiftH}), and the deceleration parameter $\mathcal{Q}$ (\ref{shiftQ}). In contrast to the approach necessary in \cite{Umeh:2010pr,Brown:2012fx}, the Hubble rate can be simply and physically associated with the expansion scalar of the cosmological fluid. $\Delta$ and $\mathcal{Q}$ are not immediately related to parameters in an averaged cosmology written in terms of effective Friedmann equations, but $\Delta$ can be straightforwardly related to an effective energy density in the backreaction and, if it were required, $\mathcal{Q}$ could be linked with an effective pressure through the simple commutation between time differentiation and spatial averaging that arises in a VPC.

Since we understand linear perturbation theory in Fourier space, the averages $\Delta$ and $\mathcal{Q}$ can be evaluated for a typical event through ensemble averaging. To do so we employed a Boltzmann code progressively developed from COSMICs and CMBFast \cite{Bertschinger:1995er,Seljak:1996} in \cite{Behrend:2008,Brown:2009a,Brown:2009cy,Brown:2012fx}, which we used to recover $\phi(k,\eta_0)$, the linear Newtonian potential at the present epoch. In principle, this allows us to evaluate $\overline{\Delta}$ and $\overline{\mathcal{Q}}$ (and if we desired their variances, which were noted in \cite{Clarkson:2009} to potentially be far more significant than the means).

However, integrands in the VPC exhibit infra-red and ultra-violet divergences which are endemic in studies of second-order perturbation theory. Both infra-red and ultra-violet divergences are familiar in studies of backreaction undertaken in 3-domains. While they occur in 3-averages evaluated in Newtonian gauge \cite{Umeh:2010pr} or traceless uniform CDM gauge \cite{Brown:2012fx}, they do not appear in uniform curvature gauge \cite{Brown:2012fx} (although it is unclear whether or not they occur in the deceleration parameter in this case). Therefore, the divergences are clearly a gauge-dependent effect. The ultra-violet divergence in $\mathcal{Q}$ that we found in this study is similar in form to that which appeared in the traceless uniform CDM gauge. We controlled the infra-red and ultra-violet divergences in the usual manner, imposing a hard large-scale cut-off at $k_{\rm IR}=k_H/10$ and damping perturbations on scales below $k_{\rm UV}$. The large-scale cut-off can be justified by noting that the divergence is only logarithmic, and that a change in $k_{\rm IR}$ by an order of magnitude only influences the results by a factor of approximately 2. On small scales, linear perturbations are damped by Silk damping and it is reasonable to take $k_{\rm UV}=k_{\rm Silk}\approx 6\rm{Mpc}^{-1}$; furthermore, on such scales linear perturbation theory becomes invalid and the results from perturbation theory can no longer be trusted.

Solving the integrals numerically, the results were presented in Figure \ref{Fig:Integrals}. We found that the fractional shift in the Hubble parameter with respect to the input FLRW model is $\overline{\Delta}\approx 10^{-5}$, which leads to $\Omega_{\rm eff}$ of the order of a few times $10^{-5}$. This is broadly consistent with previous results. However, it is interesting to compare this result with the effective energy density obtained in uniform curvature gauge, which was found to be significantly larger: $\Omega_{\rm eff}\approx 5\times 10^{-4}$. It seems plausible to suggest that this result in uniform curvature gauge provides an approximate upper-bound on the effective energy density from backreaction; since the integrands exhibited neither ultra-violet nor infra-red divergences, we did not have to neglect the impact from any scale in the manner we have had to here. By imposing large-scale cut-offs and small-scale smoothing we may be neglecting regions of the average that we should instead consider -- note, for instance, that imposing a cut-off on large-scales in Fourier space corresponds to removing concentric shells, or ``onion rings'', from a spherical average. The question of the relative size of these effects and how self-consistent averaging within perturbation theory may affect the correct interpretation of cosmological observations has been discussed previously (see for instance \cite{Li08,Larena:2008,Clarkson:2011zq,Brown:2012fx}).

The fractional shift in the averaged deceleration parameter evaluated in the VPC was $\overline{\mathcal{Q}}\approx 0.44$. The size of the fluctuation to the averaged acceleration rate is very large, being driven by the rapid ultra-violet divergence generating the large peak visible in the integrand. It is unclear to what extent this is a gauge-dependent effect and whether it is physical or not, or whether it indicates a breakdown of the formalism.

As we have emphasised, a cosmological VPC is defined by a single scalar gauge condition. In this study we set the second gauge-condition by requiring that the spatial curvature $\psi_V=0$. However, this choice was arbitrary and it seems likely that imposing an alternative gauge condition would remove one, or perhaps both, of these divergences and produce a gauge better-suited to the problem. Defining a VPC by the requirement that infra-red or ultra-violet divergences do not occur in $\overline{\mathcal{Q}}$ or $\overline{\Delta}$, could be expected to provide the ``preferred'' coordinate system (or systems, in the event that both cannot be safely evaluated in a single VPC) in which to perform spatial averaging. As a concrete example, it is possible to choose a gauge in which the backreaction on the Hubble rate in a 3-volume exhibits neither infra-red nor ultra-violet divergences \cite{Brown:2012fx}. Alternatively we could choose the second gauge condition to explicitly eliminate divergences in the averaged acceleration rate. While large-scale cut-offs and small-scale divergences control our integrands, they are unsatisfactory for two chief reasons: first, they are entirely arbitrary and the extreme sensitivity on $k_{\rm UV}$ in particular creates large changes in the results for a small change in the cut-off scale; and second, we exclude the ``onion-rings'' from the spatial averages in a distinctly unphysical manner. It can be argued that the spatial average is only properly evaluated in a coordinate system in which spurious divergences do not occur and the impact from all scales (on which the theory is valid) is well-defined. This interesting prospect is left to further study; here it suffices to note that an infra-red divergence is inevitable if the ensemble average contains a constant term in its integrand, $\int A d(\ln k)$, and that an ultra-violet divergence is inevitable if it contains a term proportional to $k^4$, $\int B(k\eta)^4 d(\ln k)$. We are then effectively restricted to integrals of the form $\int C(k\eta)^2 d(\ln k)$, as seen in uniform curvature gauge in \cite{Brown:2012fx}, and the amplitude of cosmological perturbations then implies that the fractional shifts in both $\overline{\Delta}$ and $\overline{\mathcal{Q}}$ \emph{arising from cosmological perturbations} are bounded by $\{\overline{\Delta},\overline{\mathcal{Q}}\}\lesssim 10^{-3}$. This estimate is provided for Einstein-de Sitter models; for more general cosmologies the introduction of a cosmological constant tends to reduce the effect by a factor of two or three. It should also be noted that fractional shifts of this order-of-magnitude would be extremely significant. Not only are observational quantities such as the angular diameter distance of the BAOs influenced by integral effects, which could in principle compound small shifts to produce a larger impact, but the actual backreaction in the universe arises not only from perturbations but also from non-linear structures, which are likely to contribute a larger effect.

For these reasons the further investigation of VPCs in which consistent averages can be taken is vital for a full understanding of modern cosmology, and the study in this paper provides a vital first step in this procedure.

\acknowledgments{The authors wish to thank Adam Christopherson for very helpful discussions. AAC acknowledges financial support from NSERC.}

\bibliography{Paper}

\appendix
\section{Gauge Transformations at Linear and Second Order}
\noindent Here we summarise the scalar gauge transformations presented in \cite{Malik:2008,Christopherson:2011ra}. The transformation is generated by a 4-vector
\be
\xi^\mu=(\alpha,\xi^i)=(\alpha,\pd^i\beta).
\ee

\vspace{1.5em}\noindent\textbf{Linear}\newline
\noindent A four-scalar, and specifically the density perturbation, transforms as
\be
\label{rhotransform1}
\widetilde{\delta\rho}_1=\delta\rho_1+\bkr'\alpha_1 \Rightarrow \tilde{\delta}=\delta-3(1+w)\alpha_1 .
\ee
The scalar component of a four-vector, and specifically the 4-velocity, transforms as
\be
\widetilde{v}_1=v_1-\beta_1'.
\ee
The scalar metric perturbations transform as
\be
\label{linearScalarMatricGaugeTransforms}
\widetilde{\phi}_1=\phi_1+\alpha_1'+\mH\alpha_1,\quad
\widetilde{B}_1=B_1-\alpha_1+\beta_1',\quad
\widetilde{\psi}_1=\psi_1-\mH\alpha_1,\quad
\widetilde{E}_1=E_1+\beta_1 .
\ee

\vspace{1.5em}\noindent\textbf{Second Order}\newline
\noindent Four-scalars transform as
\be
\label{deltarho2trans}
\widetilde{\delta\rho}_2=\delta\rho_2+\bkr'\alpha_2+\alpha_1\left(\bkr''\alpha_1+\bkr'\alpha_1'+2\delta\rho_1'\right)+\pd_k\left(2\delta\rho_1+\bkr'\alpha_1\right)\left(\pd^k\beta_1+\gamma_1^k\right)
\ee
and the scalar component of the 4-velocity transforms as
\be
\widetilde{v}_2=v_2-\beta_2'+\pd^{-2}\pd_k\Xv^k
\ee
where $\Xv_i$ contains the terms quadratic in the first-order perturbations and is defined in equation (\ref{Xvdef}). The scalar metric components transform as
\bea
\label{transpsi2}
\widetilde{\psi}_2&=&\psi_2-\mH\alpha_2-\frac{1}{4}\X^k_k+\frac{1}{4}\pd^{-2}\pd_i\pd_j\X^{ij}, \\
\label{transB2}
\widetilde{B}_2&=&B_2-\alpha_2+\beta_2'+\pd^{-2}\pd_k\XB^k, \\
\label{transPhi2}
\widetilde{\phi}_2&=&\phi_2+\mH\alpha_2+\alpha_2'+\alpha_1\left[\alpha_1''+5\mH\alpha_1'+\left(\mH'+2\mH^2\right)\alpha_1+4\mH\phi_1+2\phi_1'\right]
\nonumber\\&&\quad
+2\alpha_1'\left(\alpha_1'+2\phi_1\right)+\xi_{1k}\pd^k\left(\alpha_1'+\mH\alpha_1+2\phi_1\right)
+\xi_{1k}'\left[\pd^k\alpha_1-2B_1^k-{\xi_1^k}'\right], \\
\label{transE2}
\widetilde{E}_2&=&E_2+\beta_2+\frac{3}{4}\pd^{-2}\pd^{-2}\pd_i\pd_j\X^{ij}-\frac{1}{4}\pd^{-2}\X^k_k,
\eea
where $\X_{ij}$ (\ref{Xdef}) and $\X^B_{i}$ (\ref{XBdef}) contain the terms quadratic in the first order perturbations. The quadratic terms are
\bea
\label{Xdef}
\X_{ij}&\equiv&2\left[\left(\mH^2+\frac{a''}{a}\right)\alpha_1^2+\mH\left(\alpha_1\alpha_1'+\xi_1^k\pd_k\alpha_1\right)\right]\delta_{ij}
+4\left[\alpha_1\left(C_{1ij}'+2\mH C_{1ij}\right)+\xi_1^k\pd_kC_{1ij}+C_{1ik}\pd_j\xi_1^k+C_{1kj}\pd_i\xi_1^k\right]
\nonumber\\&&\quad
+2\left(B_{1i}\pd_j\alpha_1+B_{1j}\pd_i\alpha_1\right)+4\mH\alpha_1\left(\pd_j\xi_{1i}+\pd_i\xi_{1j}\right)
-2\pd_i\alpha_1\pd_j\alpha_1+2\pd_i\xi_{1k}\pd_j\xi_1^k
\nonumber\\&&\quad
+\alpha_1\left(\pd_j\xi_{1i}'+\pd_i\xi_{1j}'\right)+\left(\pd_j\pd_k\xi_{1i}+\pd_i\pd_k\xi_{1j}\right)\xi_1^k
+\pd_k\xi_{1i}\pd_j\xi_1^k+\pd_k\xi_{1j}\pd_i\xi_1^k+\xi_{1i}'\pd_j\alpha_1+\xi_{1j}'\pd_i\alpha_1,
\eea
\bea
\label{XBdef}
\XB_i&\equiv&
2\Big[\left(B_{1i}'+2\mH B_{1i}\right)\alpha_1+\xi_1^k\pd_kB_{1i}-2\phi_1\pd_i\alpha_1+B_{1k}\pd_i\xi_1^k+B_{1i}\alpha_1'+2 C_{1ik}{\xi_{1}^k}'\Big]
+4\mH\alpha_1\left(\xi_{1i}'-\pd_i\alpha_1\right)
\nonumber\\&&\quad
+\alpha_1'\left(\xi_{1i}'-3\pd_i\alpha_1\right)+\alpha_1\left(\xi_{1i}''-\pd_i\alpha_1'\right)
+{\xi_{1}^k}'\left(\pd_k\xi_{1i}+2\pd_i\xi_{1k}\right)+\xi_{1}^k\left(\pd_k\xi_{1i}'-\pd_i\pd_k\alpha_1\right)-\pd_k\alpha_1\pd_i\xi_1^k ,
\eea
and
\bea
\label{Xvdef}
\Xv_i\equiv\xi_{1i}'\left(2\phi_1+\alpha_1'+2\mH\alpha_1\right)-\alpha_1\xi_{1i}''-\xi^k_1\pd_k\xi_{1i}'+{\xi_1^k}'\pd_k\xi_{1i}
-2\alpha_1\left(v'_{1i}+\mH v_{1i}\right)+2\xi^k_1\pd_kv_{1i}-2v^k_1\pd_k\xi_{1i} .
\eea

\section{Second-Order Einstein Field Equations}
\noindent In our notation, and correcting typographical mistakes in \cite{Christopherson:2011ra}, the scalar Einstein field equations at second order in Poisson gauge \cite{Nakamura:2004rm,Christopherson:2011ra,Christopherson:2013} and assuming perfect fluid matter are
\bea
0-0&:&3\mH\left(\psi_{2N}'+\mH\phi_{2N}\right)-\pd^a\pd_a\psi_{2N}-3\psi'_{1N}\hspace{-0.65em}{}^2-3\pd^a\phi_{1N}\pd_a\phi_{1N}-8\psi_{1N}\pd^a\pd_a\psi_{1N}-12\mH^2\phi_{1N}^2
\nonumber\\&&\quad
+12\mH\psi'_{1N}\left(\psi_{1N}-\phi_{1N}\right)=-4\pi Ga^2\left(\delta\rho_{2N}+2\left(\bkr+\bkp\right)\pd^av_{1N}\pd_av_{1N}\right), \\
0-i&:&\pd_i\psi_{2N}'+\mH\pd_i\phi_{2N}-8\mH\phi_{1N}\pd_i\phi_{1N}+2\psi_{1N}'\pd_i\left(2\psi_{1N}-\phi_{1N}\right)+4\left(\psi_{1N}-\phi_{1N}\right)\pd_i\psi_{1N}'
\nonumber\\&&\quad
=4\pi Ga^2\left(2\left(\delta\rho_{1N}+\delta p_{1N}\right)\pd_iv_{1N}-\left(\bkr+\bkp\right)\left(\pd_iv_{2N}-2\phi_{1N}\pd_iv_{1N}\right)\right), \\
i-j\;\mathrm{trace}&:&\psi''_{2N}+\mH\left(2\psi'_{2N}+\phi'_{2N}\right)+\frac{1}{3}\pd^a\pd_a\left(\phi_{2N}-\psi_{2N}\right)+\left(2\mH'+\mH^2\right)\phi_{2N}-4\left(2\mH'+\mH^2\right)\phi_{1N}^2
\\ &&\quad
-\frac{7}{3}\pd^a\phi_{1N}\pd_a\phi_{1N}-\frac{8}{3}\phi_{1N}\pd^a\pd_a\phi_{1N}-8\mH\phi_{1N}\phi'_{1N}-\phi'_{1N}\hspace{-0.65em}{}^2=4\pi Ga^2\left(\frac{2}{3}\left(\bkr+\bkp\right)\pd^av_{1N}\pd_av_{1N}+\delta p_2\right), \nonumber \\
i-j\;\mathrm{traceless}&:&\pd^4\left(\psi_{2N}-\phi_{2N}\right)=12\pi Ga^2\left(\bkr+\bkp\right)\left(2\pd_a\pd_b\left(\pd^av_{1N}\pd^bv_{1N}\right)-\pd^a\pd_a\left(\pd^bv_{1N}\pd_bv_{1N}\right)\right)
\\&&\quad\quad
-14\pd^a\pd^b\phi_{1N}\pd_a\pd_b\phi_{1N}-24\pd^a\phi_{1N}\pd_a\pd^b\pd_b\phi_{1N}-2\left(\pd^a\pd_a\phi_{1N}\right)^2-8\phi_{1N}\pd^4\phi_{1N}\nonumber
\eea

\section{Second-Order Dynamics in the Comoving Volume-Preserving Coordinate System}
\label{Appendix-VPCDynamics}
\noindent The quadratic terms (\ref{Xdef}, \ref{XBdef}, \ref{Xvdef}) for the transformation between Newtonian gauge and the VPC reduce to
\bea
\XB_i&=&5\eta\left(-\phi_{1N}\partial_i\phi_{1N}+\frac{1}{4}\left(\pd_i\pd_k\phi_{1N}\pd^k\pd^{-2}\phi_{1N}+\pd_k\phi_{1N}\pd_i\pd^k\pd^{-2}\phi_{1N}\right)\right), \\
\X_{ij}&=&\left(-4\phi_{1N}^2+5\pd_k\phi_{1N}\pd^k\pd^{-2}\phi_{1N}\right)\delta_{ij}+25\pd_i\pd_k\pd^{-2}\phi_{1N}\pd_j\pd^k\pd^{-2}\phi_{1N}
\nonumber \\ && \qquad\qquad\qquad\qquad
+\frac{25}{2}\pd^k\pd^{-2}\phi_{1N}\pd_i\pd_j\pd_k\pd^{-2}\phi_{1N}-\frac{1}{2}\eta^2\pd_i\phi_{1N}\pd_j\phi_{1N}, \\
\Xv_i&=&\eta\left(\phi_{1N}\pd_i\phi_{1N}+\frac{5}{3}\left(\pd_i\pd_k\phi_{1N}\pd^k\pd^{-2}\phi_{1N}-\pd^k\phi_{1N}\pd_i\pd_k\pd^{-2}\phi_{1N}\right)\right) .
\eea
To evaluate the dynamics, expansion scalar and acceleration rate in the comoving VPC it is also useful to know the time derivatives of the generating vector, the trace and inverse Laplacian of the fully contracted gradients of the quadratic term $\X_{ij}$ and the time derivative of the Newtonian gauge curvature. These are
\bea
\alpha_2'&=&\frac{1}{2}\psi_{2N}+\frac{1}{8}\pd^{-2}\pd_i\pd_j\X^{ij}+\frac{1}{8}\eta\pd^{-2}\pd_i\pd_j{\X^{ij}}'-\frac{1}{8}\X^k_k-\frac{1}{8}\eta{\X^k_k}',
 \\
\pd^a\pd_a\beta_2'&=&-\frac{25}{21}\eta\pd^{-2}\pd_i\pd^j\left(\pd^i\phi_{1N}\pd_j\phi_{1N}\right)-\frac{1}{7}\eta\pd_i\phi_{1N}\pd^i\phi_{1N}
\nonumber\\&&\quad
-\frac{5}{4}\pd^{-2}\pd_i\pd_j{\X^{ij}}'-\frac{1}{8}\eta\pd^{-2}\pd_i\pd_j{\X^{ij}}''+\frac{3}{4}{\X^k_k}'+\frac{1}{8}\eta{\X^k_k}'',
 \\
\label{Xkk}
\X^k_k&=&-12\phi_{1N}^2+25\pd_i\pd_k\pd^{-2}\phi_{1N}\pd^i\pd^k\pd^{-2}\phi_{1N}+\frac{55}{2}\pd^k\pd^{-2}\phi_{1N}\pd_k\phi_{1N}-\frac{1}{2}\eta^2\pd_i\phi_{1N}\pd^i\phi_{1N}
 \\
\label{invLXij}
\pd^{-2}\pd_i\pd_j\X^{ij}&=&
-4\phi_{1N}^2+5\pd_k\phi_{1N}\pd^k\pd^{-2}\phi_{1N}
+25\pd^{-2}\pd^i\pd^j\left(\pd_i\pd_k\pd^{-2}\phi_{1N}\pd_j\pd^k\pd^{-2}\phi_{1N}\right)
\nonumber\\&&\quad
+\frac{25}{2}\pd^{-2}\pd^i\pd^j\left(\pd^k\pd^{-2}\phi_{1N}\pd_i\pd_j\pd_k\pd^{-2}\phi_{1N}\right)
-\frac{1}{2}\eta^2\pd^{-2}\pd^i\pd^j\left(\pd_i\phi_{1N}\pd_j\phi_{1N}\right),
 \\
\psi_{2N}'&=&\frac{1}{7}\eta\left(N-\pd^a\phi_{1N}\pd_a\phi_{1N}\right)
=\frac{1}{7}\eta\left(\frac{10}{3}\pd^{-2}(\pd_i\pd^j(\pd^i\phi_{1N}\pd_j\phi_{1N}))-\pd^a\phi_{1N}\pd_a\phi_{1N}\right) .
\eea
Applying these to the second-order metric perturbations and velocity in Poisson gauge gives the VPC quantities
\bea
\phi_{2V}&=&\phi_{2N}+\frac{3}{2}\psi_{2N}+\frac{3}{8}\pd^{-2}\pd_i\pd_j\X^{ij}+\frac{1}{8}\eta(\pd^{-2}\pd_i\pd_j\X^{ij})'
-\frac{3}{8}\X^i_i-\frac{1}{8}\eta(\X^i_i)'+\frac{21}{2}\phi_{1N}^{2}-\frac{35}{4}\pd_i\phi_{1N}\pd^i\pd^{-2}\phi_{1N},
 \\
\pd^a\pd_aB_{2V}&=&-\frac{1}{2}\eta\pd^a\pd_a\psi_{2N}
-\frac{1}{8}\eta\pd_i\pd_j\X^{ij}
-\frac{5}{4}(\pd^{-2}\pd_i\pd_j\X^{ij})'
-\frac{1}{8}\eta(\pd^{-2}\pd_i\pd_j\X^{ij})''
+\frac{1}{8}\eta\pd^a\pd_a\X^i_i
+\frac{3}{4}(\X^i_i)'
+\frac{1}{8}\eta(\X^i_i)''
\nonumber\\&&\quad
-\frac{25}{21}\eta\pd^{-2}(\pd_i\pd^j(\pd^i\phi_{1N}\pd_j\phi_{1N}))
-\frac{36}{7}\eta\pd^i\phi_{1N}\pd_i\phi_{1N}
-5\eta\phi_{1N}\pd^a\pd_a\phi_{1N}
+\frac{5}{4}\eta\Big(\pd_i\pd^k\pd^{-2}\phi_{1N}\pd_k\pd^i\phi_{1N}
\nonumber\\&&\quad
+\pd^k\pd^{-2}\phi_{1N}\pd_k\pd^a\pd_a\phi_{1N}
+\pd_i\pd_k\phi_{1N}\pd^k\pd^i\pd^{-2}\phi_{1N}
+\pd_k\phi_{1N}\pd^k\phi_{1N}
\Big), \\
\pd^a\pd_aE_{2V}&=&
-\phi_{2N}-\frac{3}{2}\psi_{2N}
-\frac{3}{8}\pd^{-2}\pd_i\pd_j\X^{ij}
+\frac{3}{8}\X^i_i
+\frac{1}{8}\eta(\pd^{-2}\pd_i\pd_j\X^{ij})'
+\frac{1}{8}\eta(\X^i_i)'
+4\phi_{1N}^2
\nonumber\\&&\quad
-\frac{1}{4}\eta^2\pd^i\phi_{1N}\pd_i\phi_{1N}
+\frac{35}{4}\pd_i\pd^{-2}\phi_{1N}\pd^i\phi_{1N}
+\frac{25}{2}\pd_i\pd_j\pd^{-2}\phi_{1N}\pd^i\pd^j\pd^{-2}\phi_{1N},
 \\
\pd^a\pd_av_{2V}&=&
-\eta\phi_{1N}\pd^a\pd_a\phi_{1N}
-\frac{33}{7}\eta\pd^i\phi_{1N}\pd_i\phi_{1N}
+\frac{67}{21}\eta\pd^{-2}\left(\pd_i\pd^j\left(\pd^i\phi_{1N}\pd_j\phi_{1N}\right)\right)
\nonumber\\&&\quad
+\frac{1}{9}\eta^3\left[\left(\pd^a\pd_a\phi_{1N}\right)^2
+\pd^i\phi_{1N}\pd_i\pd^a\pd_a\phi_{1N}
-\frac{3}{7}\pd^a\pd_a\left(N-\pd^b\phi_{1N}\pd_b\phi_{1N}\right)\right]
\nonumber\\&&\quad
+\eta\left(\frac{5}{3}\pd^a\pd_a\pd_k\phi_{1N}\pd^k\pd^{-2}\phi_{1N}
+\frac{5}{3}\pd^i\pd_k\phi_{1N}\pd_i\pd^k\pd^{-2}\phi_{1N}
-\frac{5}{3}\pd_i\pd^k\phi_{1N}\pd^i\pd_k\pd^{-2}\phi_{1N}
-\frac{5}{3}\pd^k\phi_{1N}\pd_k\phi_{1N}\right)
\nonumber\\&&\quad
+\frac{5}{4}\pd^{-2}\pd_i\pd_j{\X^{ij}}'
+\frac{1}{8}\eta\pd^{-2}\pd_i\pd_j{\X^{ij}}''
-\frac{3}{4}{\X^k_k}'
-\frac{1}{8}\eta{\X^k_k}'' .
\eea

\section{Spatial Averages in Fourier space}
\label{Appendix-FourierSpace}
\noindent In this appendix we demonstrate averages of expressions of the form $\pd^{-2}\pd_i\pd_jC^{ij}$, where $C^{ij}=\pd^iA\pd^jB$ or $C^{ij}=A\pd^i\pd^jB$. The starting point is equation (\ref{AverageConvolution}). Then the average of both of these terms can be written
\be
\overline{\av{X(\xv)}}_3=-\frac{1}{V_3}\iint W^*(\kv)\frac{f(\kv,\kv')}{k^2}\overline{A(\kv')B(\kv-\kv')}\frac{d^3\kv'}{(2\pi)^3}\dtk
\ee
where
\be
f(\kv,\kv')=
\left\{\begin{array}{rl}
k^2(kk'\cos\theta_{k'}-{k'}^2\cos^2\theta_{k'})=k^2g_1(\kv,\kv'),& C^{ij}=\pd^iA\pd^jB\\
k^2\left(k^2+{k'}^2\cos^2\theta_{k'}-2kk'\cos\theta_{k'}\right)=k^2g_2(\kv,\kv'),& C^{ij}=A\pd^i\pd^jB
\end{array}\right.
\ee
which can be found by expanding $A(\xv)$ and $B(\xv)$ across the Fourier modes and evaluating the derivatives. The power spectrum for $A$ and $B$ then gives
\be
\overline{\av{X(\kv)}}_3=-\frac{1}{V_3}\frac{1}{4\pi}\iint W^*(\kv)\frac{g(\kv,\kv')}{k'^3}A(k')B(k')\mathcal{P}(k')\delta(\kv)d^3\kv'd^3\kv .
\ee
We can now perform the integral over $\theta_{k'}$ and $\phi_{k'}$, giving
\be
\overline{\av{X(\kv)}}_3=-\frac{1}{2V_3}\iint W^*(\kv)I(k,k')A(k')B(k')\mathcal{P}(k')\delta(\kv)\frac{dk'}{k'}d^3\kv .
\ee
with
\be
I(k,k')=
\left\{\begin{array}{rl}
-\dfrac{2}{3}{k'}^2,& C^{ij}=\pd^iA\pd^jB\\
\left(2k^2+\dfrac{2}{3}k'^2\right),& C^{ij}=A\pd^i\pd^jB
\end{array}\right. .
\ee
The integral over $\kv$ is now trivial. Noting that the term proportional to $k^2$ in $I(k,k')$ vanishes in the integral, we are left with
\be
\overline{\av{X(\kv)}}_3=\pm\frac{1}{3}\int k'^2\mathcal{P}(k')A(k')B(k')\frac{dk'}{k'}
\ee
where the negative sign is taken for $C^{ij}=\pd^iA\pd^jB$ while the positive sign is taken for $C^{ij}=A\pd^i\pd^jB$. The result for $C^{ij}=\pd^iA\pd^jB$ agrees with that in \cite{Clarkson:2009} and to our knowledge the other is original to this paper.

\end{document}